\documentclass{pasj00}

\begin{document}
\SetRunningHead{M. Oguri}{SDSS J1004+4112 Revisited}
\Received{2010/XX/XX}
\Accepted{2010/XX/XX}

\title{
The Mass Distribution of SDSS J1004+4112 Revisited} 

\author{
Masamune \textsc{Oguri}\altaffilmark{1,2}
}
\altaffiltext{1}{Division of Theoretical Astronomy, National Astronomical
Observatory of Japan, 2-21-1 Osawa, Mitaka, Tokyo 181-8588, Japan.}
\altaffiltext{2}{Kavli Institute for Particle Astrophysics and
Cosmology, Stanford University, 2575 Sand Hill Road, Menlo Park, CA
94025, USA.} 


\maketitle

\begin{abstract}
We present a strong lens analysis of SDSS~J1004+4112, a unique quasar
lens produced by a massive cluster of galaxies at $z=0.68$, using a
newly developed software for gravitational lensing. We find that our
parametric mass model well reproduces all observations including the
positions of quasar images as well as those of multiply imaged
galaxies with measured spectroscopic redshifts, time delays between
quasar images, and the positions of faint central images. The
predicted large total magnification of $\mu \sim 70$ suggests that the
lens system is indeed a useful site for studying the fine structure of
a distant quasar and its host galaxy. The dark halo component is found
to be unimodal centered on the brightest cluster galaxy and the {\it
  Chandra} X-ray surface brightness profile. In addition, the
orientation of the halo component is quite consistent with those of
the brightest cluster galaxy and member galaxy distribution, implying
that the lensing cluster is a relaxed system. The radial profile of
the best-fit mass model is in good agreement with a mass profile
inferred from the X-ray observation. While the inner radial slope of
the dark halo component is consistent with being $-1$, a clear
dependence of the predicted A--D time delay on the slope indicates
that an additional time delay measurement will improve constraints on
the mass model. 
\end{abstract}

\section{Introduction}

SDSS~J1004+4112 is a unique quasar lens system 
\citep{inada03,oguri04,inada05,inada08}. A quasar at $z=1.734$ is
multiply imaged into five images, with their maximum image separation
of $14\farcs7$, produced by a massive cluster of galaxies at
$z=0.68$. It is one of only two known examples of cluster-scale quasar
lenses, the other being the triple lens SDSS~J1029+2623 with the
maximum image separation of $22\farcs5$ \citep{inada06,oguri08}. In
addition to the quasar images, SDSS~J1004+4112 contains
spectroscopically confirmed multiply imaged galaxies at $z\sim 3$
\citep{sharon05}. Both the quasar images and the lensing cluster have
been detected in {\it Chandra} X-ray observations \citep{ota06}. 
Moreover, time delays between some of the quasar images have also been
measured from long-term optical monitoring observations
\citep{fohlmeister07,fohlmeister08}. 

\begin{table*}
  \caption{Previous mass modeling of SDSS~J1004+4112\label{table:model}}
  \begin{center}
    \begin{tabular}{lccc}
     \hline\hline
      Reference & model\footnotemark[$*$] & constraints (quasar)\footnotemark[$\dagger$] & constraints (galaxy)\footnotemark[$\dagger$]\\
     \hline
 \citet{inada03}   & SIE+pert      & pos+flux (4) & $\cdots$ \\
 \citet{oguri04}   & NFW+SIE+pert  & pos+flux (4) & $\cdots$ \\
 \citet{williams04}& non-parametric& pos (4)      & $\cdots$ \\
 \citet{sharon05}  & SPL+gals      & pos (5)      & pos (5)  \\
 \citet{kawano06}  & gNFW+SIE+pert & pos+flux (4) & $\cdots$ \\
 \citet{saha06}    & non-parametric& pos (5)      & pos (8) \\
 \citet{fohlmeister07} & NFW+deV+gals+pert & pos+flux (5) & $\cdots$ \\
 \citet{saha07}    & non-parametric& pos (5), $\Delta t$ (2) & pos (8) \\
 \citet{inada08}   & NFW+SIE+gals+pert & pos+flux (5), $\Delta t$ (2) & $\cdots$\\
 \citet{liesenborgs09}& non-parametric & pos (5), $\Delta t$ (3) & pos (7) \\
\hline
 This work & gNFW+Jaffe+gals+pert & pos+flux (5), $\Delta t$ (3) & pos (27)\\
   \hline
     \multicolumn{4}{@{}l@{}}{\hbox to 0pt{\parbox{160mm}{\footnotesize
       \footnotemark[$*$] Name of models: SIE = singular isothermal
       ellipsoid, pert = external perturbation (e.g., external shear),
       NFW = Navarro, Frenk \& White (NFW) profile, SPL = softened
       power-law model, gals = perturbations from member galaxies,
       typically modeled by truncated isothermal profiles, gNFW =
       generalized NFW profile, dev = de Vaucouleurs profile, Jaffe =
       pseudo-Jaffe profile. 
       \par\noindent
       \footnotemark[$\dagger$] ``pos'' indicates constraints from
       image positions, ``flux'' from fluxes of images, and $\Delta t$
       from time delays. Numbers in parentheses show the number of
       images used as constraints. }\hss}}
\end{tabular}
  \end{center}
\end{table*}

Such a wealth of observational data available enable detailed
investigations of the central mass distribution of the lensing
cluster. Indeed, there have been several attempts to model the mass
distribution of SDSS~J1004+4112, using either parametric or
non-parametric method. \citet{oguri04} adopted two-component (halo
plus central galaxy) model to successfully reproduce the positions of
four quasar images, but even the parities and temporal ordering of the
quasar images could not be determined because of model degeneracies. 
\citet{kawano06} extended mass modeling along this line, and explored
how time delay measurements can distinguish different mass profiles. 
\citet{fohlmeister07} pointed out that it is important to include
perturbations from member galaxies to reproduce the observed time
delay between quasar images A and B. On the other hand,
\citet{williams04} and \citet{saha07} performed non-parametric mass
modeling to show possible substructures in the lensing cluster. 
From the non-parametric mass modeling, \citet{saha06} and
\citet{liesenborgs09} concluded that the radial mass profile is
consistent with that predicted in $N$-body simulation 
\citep{navarro97}. We summarize previous mass modeling in 
Table~\ref{table:model}. 

In this paper, we revisit strong lens modeling of SDSS~J1004+4112
adopting a parametric mass model. We include many observational
constrains currently available, including central images of lensed
quasars and galaxies, flux rations, and time delay between quasar
images (see Table~\ref{table:model}). In particular this paper
represents first parametric mass modeling that includes {\it both} the
quasar time delays and the positions of multiply imaged galaxies as
constraints. We compare our best-fit mass model with the {\it Chandra}
X-ray observation of this system \citep{ota06}. 

This paper is organized as follow. We describe our mass model in
\S\ref{sec:model}. We show our results in \S\ref{sec:result}, and give
conclusion in \S\ref{sec:model}. A new software for gravitational
lensing, which is used for the mass modeling, is presented in
Appendix. Throughout the paper we adopt the
matter density of $\Omega_M=0.26$ and the cosmological constant of
$\Omega_\Lambda=0.74$, but regard the dimensionless Hubble constant
$h$ as a parameter. With this choice of cosmological parameters, 
a physical transverse distance of $1h^{-1}$~kpc at the redshift of
the lensing cluster ($z=0.68$) corresponds to $0.20$~arcsec. We denote
a angular diameter distance from observer to lens as $D_{\rm l}$,
from observer to source as $D_{\rm s}$, and from lens to source as 
$D_{\rm ls}$. 

\section{Mass Modeling}\label{sec:model}

\subsection{A Parametric Model}

We model a main halo of the lensing cluster as the generalized
NFW profile (e.g., \cite{jing00}). Its radial density profile is given
by 
\begin{equation}
\rho(r)=\frac{\rho_s}{(r/r_s)^{\alpha}(1+r/r_s)^{3-\alpha}}.
\end{equation}
In this model, the inner slope is parametrized by $\alpha$
($0<\alpha<2$); the original NFW profile corresponds to $\alpha=1$.  
We adopt the following modified concentration parameter as a model
parameter: 
\begin{equation}
c_{-2}=\frac{r_{\rm vir}}{r_{-2}}=\frac{r_{\rm vir}}{(2-\alpha)r_s},
\end{equation}
where $r_{-2}$ indicates the radius where the radial slope becomes
$d\ln\rho/d\ln r=-2$ and $r_{\rm vir}$ is the virial radius of the
cluster. The characteristic density is described as
\begin{equation}
\rho_s=\frac{\Delta(z)\bar{\rho}(z)c^3}{3m_{\rm gnfw}(c)},
\end{equation}
\begin{equation}
m_{\rm gnfw}(c)=\int_0^c \frac{r^{2-\alpha}}{(1+r)^{3-\alpha}}dr,
\end{equation}
where $\Delta(z)$ is nonlinear overdensity at redshift $z$ which we
adopt values predicted by the spherical collapse model. The lensing
deflection angle is related with the projected two-dimensional mass
distribution (i.e., convergence $\kappa$) computed by
\begin{equation}
\kappa(r)=\frac{1}{\Sigma_{\rm crit}}\int_{-\infty}^\infty \rho(\sqrt{r^2+z^2})dz,
\end{equation}
with $\Sigma_{\rm crit}=(c^2/4\pi G)(D_{\rm s}/D_{\rm l}D_{\rm ls})$
being the critical surface mass density computed from angular diameter
distances between observer, lens, and source. 

We then introduce an ellipticity in the projected mass distribution
by replacing $r$ in $\kappa(r)$ by the following quantity: 
\begin{equation}
\kappa(r):\;\;\;
r\rightarrow v\equiv\sqrt{\frac{\tilde{x}^2}{(1-e)}+(1-e)\tilde{y}^2},
\label{eq:elldef}
\end{equation}
where $e$ is an ellipticity (the axis ratio is $1-e$), and
$\tilde{x}$ and $\tilde{y}$ are defined by 
\begin{eqnarray}
\tilde{x}&=&x\cos\theta_e +y\sin\theta_e,\label{eq:tilx}\\
\tilde{y}&=&-x\sin\theta_e +y\cos\theta_e.\label{eq:tily}
\end{eqnarray}
In this paper we take the $x$-axis to West and the $y$-axis to North,
and therefore the position angle $\theta_e$ is measured East of North.

We include the brightest cluster galaxy (BCG) G1 modeled by
pseudo-Jaffe Ellipsoid \citep{keeton01b}. The convergence takes the
following form:  
\begin{equation}
\kappa=\kappa_0\left[\frac{1}{\sqrt{s^2/q+\tilde{x}^2+\tilde{y}^2/q^2}}
-\frac{1}{\sqrt{a^2/q+\tilde{x}^2+\tilde{y}^2/q^2}}\right],
\label{eq:jaffe}
\end{equation}
\begin{equation}
\kappa_0=\frac{\sigma^2}{2G\Sigma_{\rm crit}D_{\rm l}^2\sqrt{q}},
\end{equation}
where $q=1-e$ is the axis ratio, $s$ is the core radius, $a$ is the
truncation radius, and $\sigma$ is the velocity dispersion. This
profile has a constant density at $r\ll s$, an isothermal density
profile $r^{-2}$ at $s\ll r\ll a$, and a sharply decline profile
$r^{-4}$ at $r\gg a$. 

Perturbations from member galaxies are also included. We model
individual galaxies by the pseudo-Jaffe ellipsoid without core radius
(eq. [\ref{eq:jaffe}] with $s=0$). We fix positions, relative
luminosities, ellipticities and position angles of member galaxies to
observed values, and assumes that the velocity dispersions and
truncation radii scale with the luminosity: 
\begin{equation}
 \frac{\sigma}{\sigma_*}=\left(\frac{L}{L_*}\right)^{1/4},
\end{equation}
\begin{equation}
 \frac{a}{a_*}=\left(\frac{L}{L_*}\right)^{1/2}.
\end{equation}
The mass-to-light ratio becomes constant with this scaling. We include
14 member galaxies within $\lesssim 20''$ from the center, which are
selected from the $gri$-band Subaru Suprime-cam images \citep{oguri04}.
We adopt $r$-band luminosities for the scaling. 

To achieve better fit, we also include additional several perturbations.
We consider general perturbations whose lens potentials $\phi$ are
described as (see, e.g., \cite{evans03}; \cite{kawano04};
\cite{congdon05}; \cite{yoo06})
\begin{equation}
\phi= -\frac{\epsilon}{m}r^2\cos m(\theta-\theta_\epsilon-\pi/2).
\end{equation}
In this paper we include four perturbation terms with $m=2$ (external
shear; e.g., \cite{keeton97}),  $3$, $4$, and $5$.

\begin{figure*}
  \begin{center}
    \FigureFile(85mm,85mm){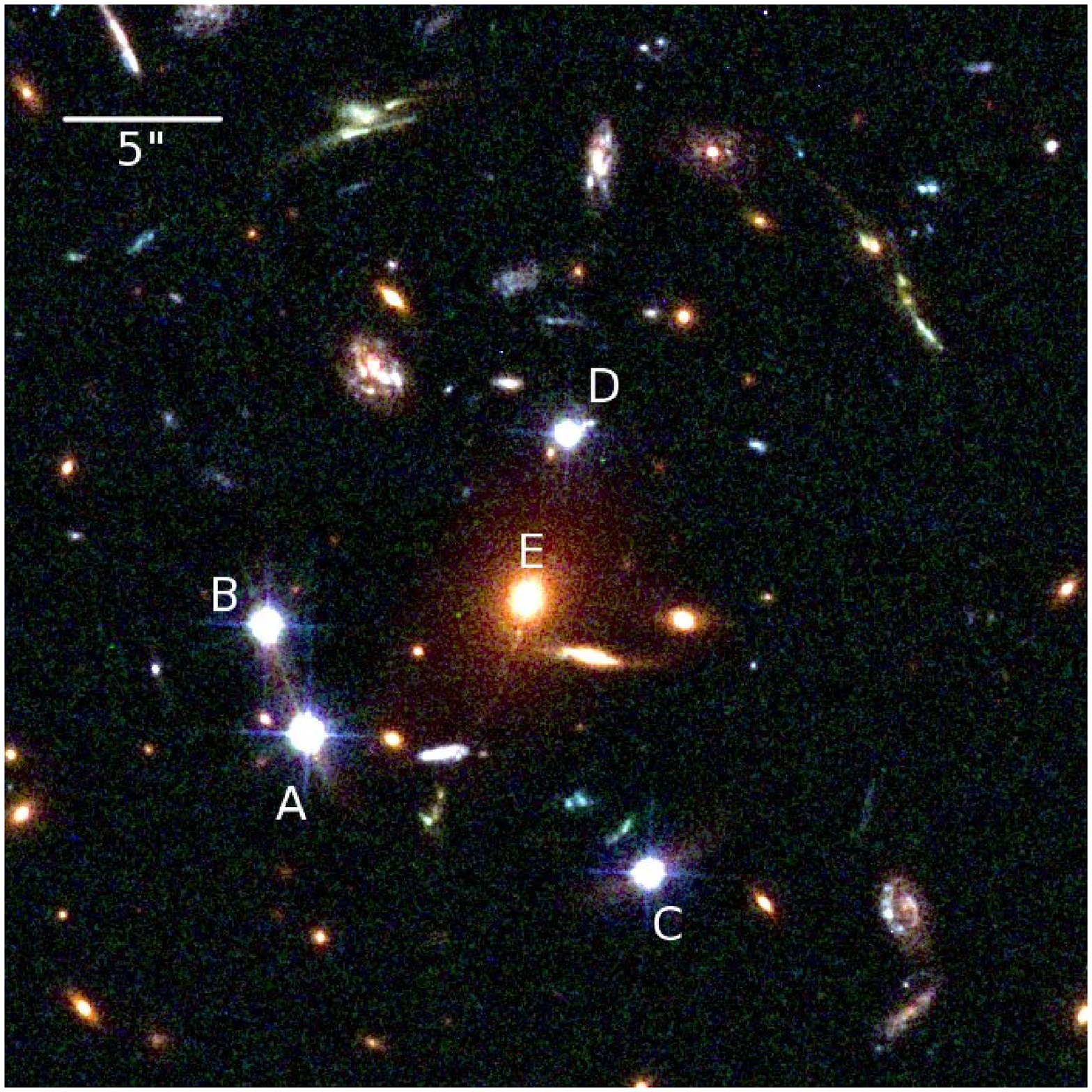}
    \FigureFile(85mm,85mm){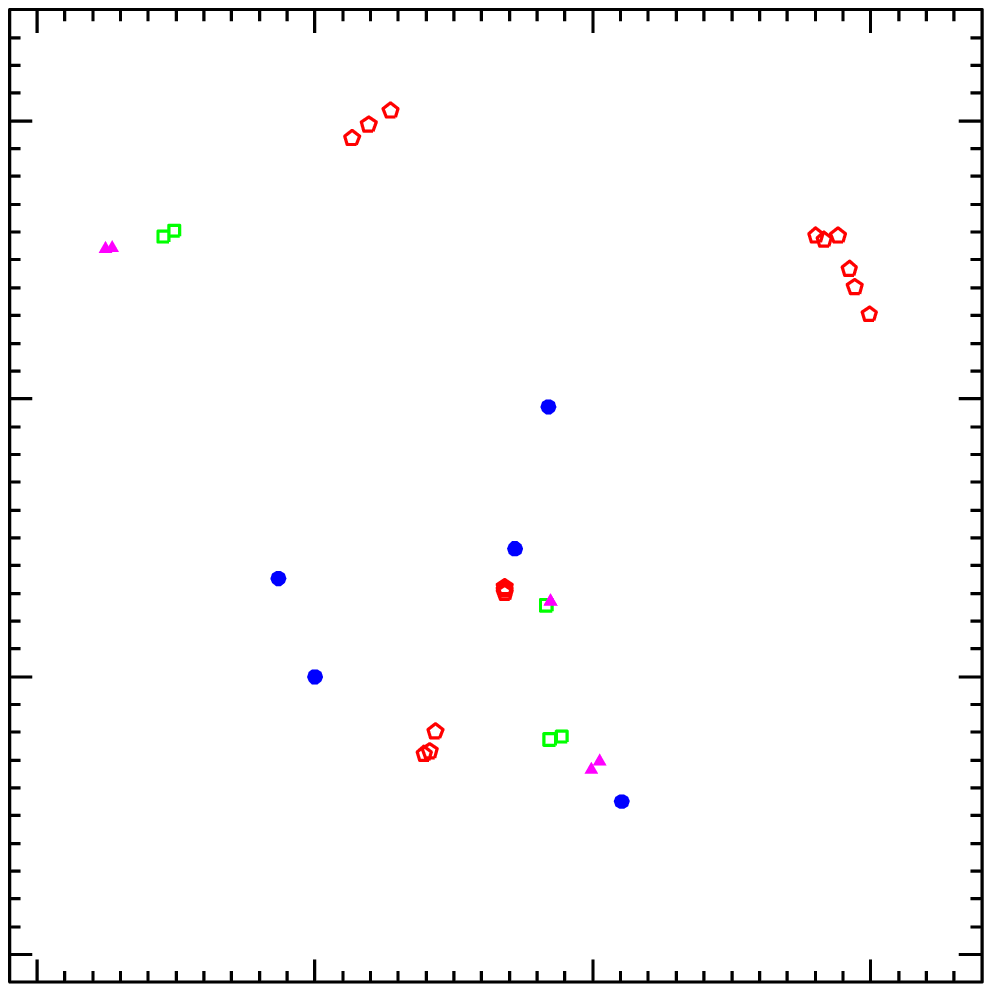}
  \end{center}
  \caption{
  {\it Left:} The HST/ACS image of SDSS~J1004+4112. North is up
  and West is right. Stellar objects labeled by A--E are the 5 lensed
  quasar images. A large galaxy superposed on image E is the brightest
  cluster galaxy G1.  
  {\it Right:} Positions of the 5 lensed quasar images ({\it filled
    circles}) and lensed galaxies images ({\it other symbols}). 
  The size and orientation of the panel is same as the left
  panel. Different symbols have different redshifts. See Tables
  \ref{table:quasar} and \ref{table:galaxy} for the relative
  coordinate values. 
  }
\label{fig:hst}
\end{figure*}

\subsection{Observational Constraints}

We adopt positions of five quasar images measured by \citet{inada05}
using the Hubble Space Telescope Advanced Camera for Surveys (HST/ACS)
F814W image. Considering possible effects of microlensing or
small-scale structure, we adopt conservative positional errors of
$0\farcs04$, and also relative magnitudes of $0.3$ ($0.8$) for image
B-D (E). In addition, we include measured time delays between image A
and B \citep{fohlmeister07} and between image A and C
\citep{fohlmeister08}. When fitting the time delays, we allow the
Hubble constant to vary with a Gaussian prior of $h=0.72\pm0.04$.  

\begin{table}
  \caption{Constraints from lensed quasar images\label{table:quasar}}
  \begin{center}
    \begin{tabular}{ccccc}
     \hline\hline
      Name & $\Delta x$ [$''$] & $\Delta y $ [$''$] & $\Delta m $ &
      $\Delta t$ [days]\\
     \hline
A & 0.000  &  0.000  &  $\equiv  0$ & $\equiv 0$  \\
B &$-1.317$&  3.532  &  $0.35\pm 0.3$ & $-40.6\pm1.8$ \\
C & 11.039 & $-4.492$&  $0.87\pm 0.3$ & $-821.6\pm2.1$\\
D &  8.399 &  9.707  &  $1.50\pm 0.3$ &  $\cdots$     \\
E &  7.197 &  4.603  &  $6.30\pm 0.8$ &  $\cdots$     \\
   \hline
\multicolumn{5}{@{}l@{}}{\hbox to 0pt{\parbox{85mm}{\footnotesize
    The quasar redshift is $z_s=1.734$. The positional error is assumed
    to $0\farcs04$ for all the lensed quasar images.}\hss}}
\end{tabular}
  \end{center}
\end{table}

\begin{table}
  \caption{Constraints from lensed galaxy images\label{table:galaxy}}
  \begin{center}
    \begin{tabular}{ccccc}
     \hline\hline
      Name & $z_s$ &$\Delta x$ [$''$] & $\Delta y $ [$''$] \\
     \hline
A1.1 & 3.33 & 3.93  & $-2.78$ \\
A1.2 &      & 1.33  & 19.37  \\
A1.3 &      &19.23  & 14.67  \\
A1.4 &      &18.83  & 15.87  \\
A1.5 &      & 6.83  &  3.22  \\
A2.1 & 3.33 &4.13   &$-2.68$  \\
A2.2 &      &   1.93&  19.87  \\
A2.3 &      &  19.43&  14.02  \\
A2.4 &      &  18.33&  15.72  \\
A2.5 &      &   6.83&   3.12  \\
A3.1 & 3.33 & 4.33  & $-1.98$ \\
A3.2 &      & 2.73  & 20.37  \\
A3.3 &      &19.95  & 13.04  \\
A3.4 &      &18.03  & 15.87  \\
A3.5 &      & 6.83  &  3.02  \\
B1.1 & 2.73 & 8.88  & $-2.16$ \\
B1.2 &      &$-5.45$& 15.84 \\
B1.3 &      & 8.33  &  2.57 \\
B2.1 & 2.73 & 8.45  & $-2.26$\\
B2.2 &      &$-5.07$& 16.04 \\
B2.3 &      &  8.33 &  2.57 \\
C1.1 & 3.28 & 10.25 &$-3.06$ \\
C1.2 &      &$-7.55$& 15.39 \\
C1.3 &      &  8.49 &  2.72 \\
C2.1 & 3.28 & 9.95  & $-3.36$\\
C2.2 &      &$-7.30$& 15.44 \\
C2.3 &      &  8.49 &  2.72 \\
   \hline
\multicolumn{4}{@{}l@{}}{\hbox to 0pt{\parbox{50mm}{\footnotesize
   The positional error is assumed to $0\farcs4$ for all the lensed
   galaxy images. The redshifts are measured spectroscopically.}\hss}} 
\end{tabular}
  \end{center}
\end{table}

We also include multiply imaged galaxies, identified by \citet{sharon05}, 
as constraints. We revisit deep multi-band HST/ACS images (F435W,
F555W, and F814W; GO-10509 and GO-10716), and identify several
features associated to each lensed images. We use positions of all
these features for our mass modeling. We include central images as
well \citep{liesenborgs09}. We assume larger positional errors of
$0\farcs4$ than those of the quasar images, partly because the
determination of the centroids of the extended galaxy images are much 
less accurate.   
   
Figure~\ref{fig:hst} shows the HST/ACS image of SDSS~J1004+4112,
together with the positions of multiple images summarized in
Tables~\ref{table:quasar} and \ref{table:galaxy}. A notable feature of
this cluster strong lens system, which can easily be seen in the
Figure, is that multiple images are distributed in a very wide range
in radius, ranging from the central images very near the cluster
center to lensed galaxy images at $\sim 30''$ from the cluster center. 
This is apparently good for constraining the density profile of the
lensing cluster.

We also add several Gaussian priors to the mass model. Based on the
measurement by \citet{inada08}, we assume the velocity dispersion of
the central galaxy G1 to $\sigma=352\pm13$~${\rm
  km\,s^{-1}}$.\footnote{Strictly speaking, the velocity dispersion
  $\sigma_{\rm obs}$ computed from the density profile can in
  principle differ from the input parameter $\sigma$ for the pseudo
  Jaffe profile. However, from \citet{eliasdottir07} we find that
  $\sigma\approx\sigma_{\rm obs}$ for values similar to those in the
  best-fit model ($s/a\approx 0.05$), which suggests that our
  assumption of $\sigma=\sigma_{\rm obs}$ is reasonable. }
The position of G1 is fixed to the observed position, ($7\farcs114$,
$4\farcs409$) in our coordinate system whose origin is at the quasar
image A. From the observed position and shape, we assume the
ellipticity and the position angle to $0.3\pm0.05$ and $152\pm 5$~deg,
respectively. Furthermore we add a weak prior to the truncation
radius, $a=8''\pm4''$, based on the observed correlation between the
velocity dispersion and the truncation radius \citep{natarajan09}.

\subsection{Model Optimization}

We use a software named {\it glafic} for all the calculations of lens
properties and model optimizations (see Appendix~\ref{sec:glafic}). 
We employ a standard $\chi^2$ minimization to find the best-fit mass
model. Specifically we adopt a downhill simplex method to find a
minimum. To speed up the calculations, we estimate $\chi^2$ in the
source plane, which is found to be sufficiently accurate for our
purpose. See Appendix~\ref{sec:schi} for detailed discussions about
the source plane $\chi^2$ minimization.

\section{Result}\label{sec:result}

\begin{figure}
  \begin{center}
    \FigureFile(85mm,85mm){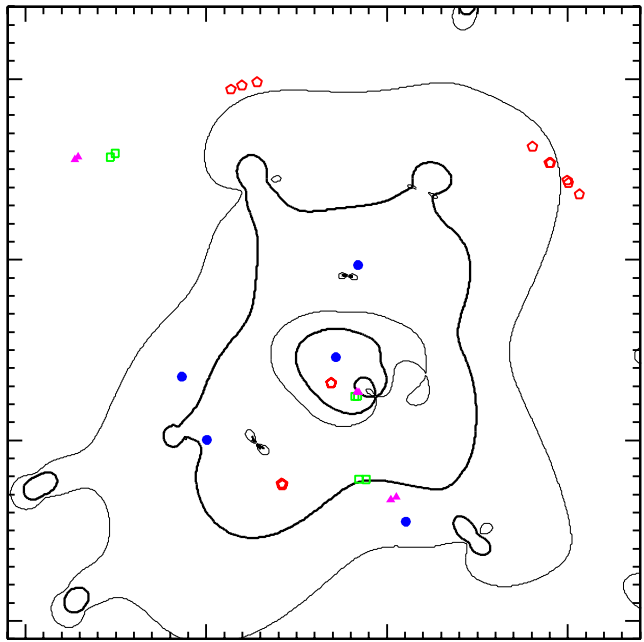}
  \end{center}
  \caption{
The best-fit mass model assuming the NFW ($\alpha=1$) profile for the
dark halo component. Symbols are same as those in the right panel of
Figure~\ref{fig:hst}, but for showing the image positions predicted by
the best-fit model. Thick solid lines indicate critical curves for the
quasar source redshift, $z_s=1.734$, whereas thin lines are critical
curves for the highest redshift among the multiply imaged galaxies,
$z_s=3.33$. 
  }
\label{fig:pos_best}
\end{figure}

\subsection{Best-fit NFW Model}

First, we fix the inner slope of the dark halo component to $\alpha=1$
(i.e., the NFW profile) and derive the best-fit mass model. 
Figure~\ref{fig:pos_best} shows best-fit critical curves. It is seen
that the best-fit model reproduces the observed multiple image
families quite well. In addition, it successfully reproduces observed
time delays between quasar images. The best-fit model has $\chi^2=31$
for 39 degree of freedom, suggesting that our choice of errors was 
reasonable. The contribution from each source is reasonably similar, 
$\chi^2=4.7$ from the quasar, $21$ from the galaxy A, $0.9$ from
galaxy B, and $2.6$ from galaxy C. The best-fit model predicts
magnifications of the five 
quasar images to $\mu_{\rm A}=29.7$, $\mu_{\rm B}=19.6$,  $\mu_{\rm
  C}=11.6$, $\mu_{\rm D}=5.8$, and $\mu_{\rm E}=0.16$. The total
magnification for all the quasar images is $\mu_{\rm tot}=67$. 
The model also predicts the time delay between quasar image A and D to
be $\Delta t_{\rm AD}=\Delta t_{\rm D}-\Delta t_{\rm A}=1218$~days,
and that between quasar image A and E to be $\Delta t_{\rm AE}=\Delta
t_{\rm   E}-\Delta t_{\rm A}=1674$~days. The AD time delay is slightly
smaller than the lower limit reported by \citet{fohlmeister08},
$\Delta t_{\rm AD}>1250$~days. 

We find that the best-fit centroid of the dark halo (NFW) component is
($6.92^{+0.20}_{-0.32}$, $4.25^{+0.31}_{-0.24}$) at 95\% confidence
limit, which is quite consistent with the observed position of G1. The
result is in marked contrast with 
\citet{oguri04}, in which significant offsets between the center of
the dark halo component and that of G1 have been reported based on
modeling of quasar images A--D. Such large offset is no longer
allowed because of additional constraints from multiply imaged
galaxies, time delays, and central images. The best-fit center of the
dark halo component is also consistent with the X-ray center reported
by \citet{ota06}. Furthermore, the best-fit position angle of
$\theta_e=152.9$~deg for the dark halo component is quite consistent
with the position angle of G1, and also that of the member galaxy
distributions studied in \citet{oguri04}. The concentricity and
alignment between dark matter, BCG, and X-ray implies that the lensing 
cluster is a highly relaxed system (see also \cite{liesenborgs09}).
The best-fit ellipticity of the halo component is $e=0.26$.

The best-fit parameters for perturbations terms are ($\epsilon$,
$\theta_\epsilon$)=(0.040, 51.8) for $m=2$, (0.019, 114) for $m=3$,
(0.013, 47) for $m=4$, and (0.010, 16.5) for $m=5$. Thus the
perturbations are rather small, but they are still important for
accurate reproduction of lensed images, particularly for those of the
quasar (see also \cite{oguri04}).

One of the most important quantity to characterize strong lensing
system is the Einstein radius $r_{\rm Ein}$. We compute the Einstein
radii $r_{\rm Ein}$ for our best-fit mass model using the following
relation:  
\begin{equation}
M(<r_{\rm Ein})=\pi r_{\rm Ein}^2\Sigma_{\rm crit}.
\end{equation}
We find $r_{\rm Ein}=8\farcs14$ for the quasar redshift $z_s=1.734$,
and $13\farcs38$ for the redshift of the lensed galaxy A, $z_s=3.33$.
If we compute $r_{\rm Ein}$ only from the dark halo component
excluding any contributions from galaxies, we obtain 
$r_{\rm Ein}=4\farcs84$ for $z_s=1.734$ and $10\farcs31$ for
$z_s=3.33$, which are quite different from those compute from the
total mass distribution. This suggests that the effect of the BCG G1
on the lens system is quite significant.

\begin{figure}
  \begin{center}
    \FigureFile(88mm,88mm){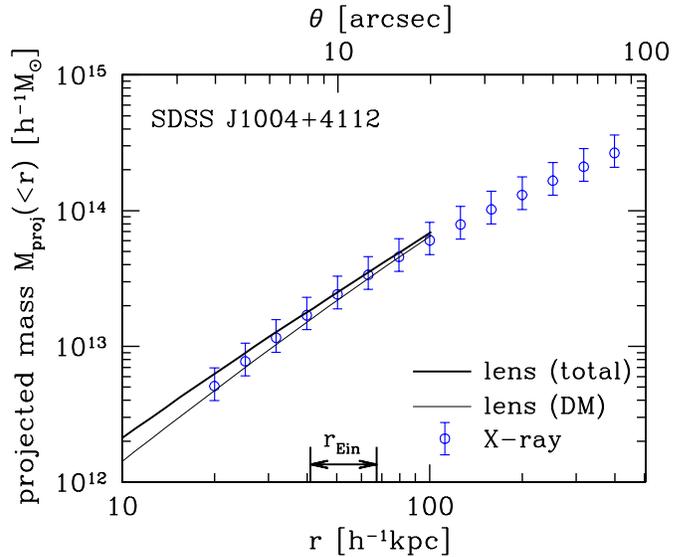}
  \end{center}
  \caption{
  Projected two-dimensional cumulative mass distributions from lensing
  and X-ray. Thick and thin lines are best-fit total and dark matter
  mass distributions from strong lens mass modeling,   respectively. 
  Open circles are mass distributions inferred from X-ray surface
  brightness and temperature measurements by {\it Chandra}, assuming
  an isothermal gas, $\beta$ model for the gas profile, and
  hydrostatic equilibrium \citep{ota06}. Errors are from the
  temperature measurement uncertainty.  
    }
\label{fig:mproj}
\end{figure}

\subsection{Comparison with X-ray}
\label{sec:xray}

Next we compare the best-fit radial mass profile derived from strong
lensing with that inferred from the {\it Chandra} X-ray observation
\citep{ota06}. In brief, from the {\it Chandra} observation the
extended X-ray emission from the lensing cluster was detected out to
$\sim 1\farcm5$ from the cluster center, with the temperature of $\sim
6.4$~keV. Assuming the isothermal profiles, \citet{ota06} constrained
the projected mass profile and argued that the mass within 100~kpc
agrees well with the mass expected from strong lensing. Here we
compare our result of new mass modeling with the X-ray result.  

Figure~\ref{fig:mproj} compares the projected two-dimensional mass
profiles from gravitational lensing and X-ray measurements. We confirm
that the profiles agree quite well with each other, including radial
slopes of the profiles. The agreement suggests that the effect of the
halo triaxiality, which affects the apparent two-dimensional lensing
masses particularly near the center of the cluster (see, e.g.,
\cite{oguri05}; \cite{gavazzi05}), is not significant.

However, it should be noted that the best-fit halo mass of $M_{\rm
  vir}=1.0\times 10^{15}h^{-1}M_\odot$ and the concentration of 
$c_{-2}=2.8$ are quite different from those inferred from X-ray, 
$M_{\rm vir}\sim 4.3\times 10^{14}h^{-1}M_\odot$ and $c_{-2}\sim
6.1$. One reason for this the strong degeneracy between $M_{\rm vir}$
and $c_{-2}$ inherent to strong lens mass modeling. Basically strong
lenses constrain the central core mass of the cluster, which is a
strong function of both $M_{\rm vir}$ and $c_{-2}$. The determination 
of $M_{\rm vir}$ and $c_{-2}$ solely from strong lensing relies on the
extrapolation of the subtle change of the radial slope out to much
larger radii. The robust determination of these parameters from
lensing therefore requires addition constraints from weak
gravitational lensing. 

\begin{figure}
  \begin{center}
    \FigureFile(85mm,85mm){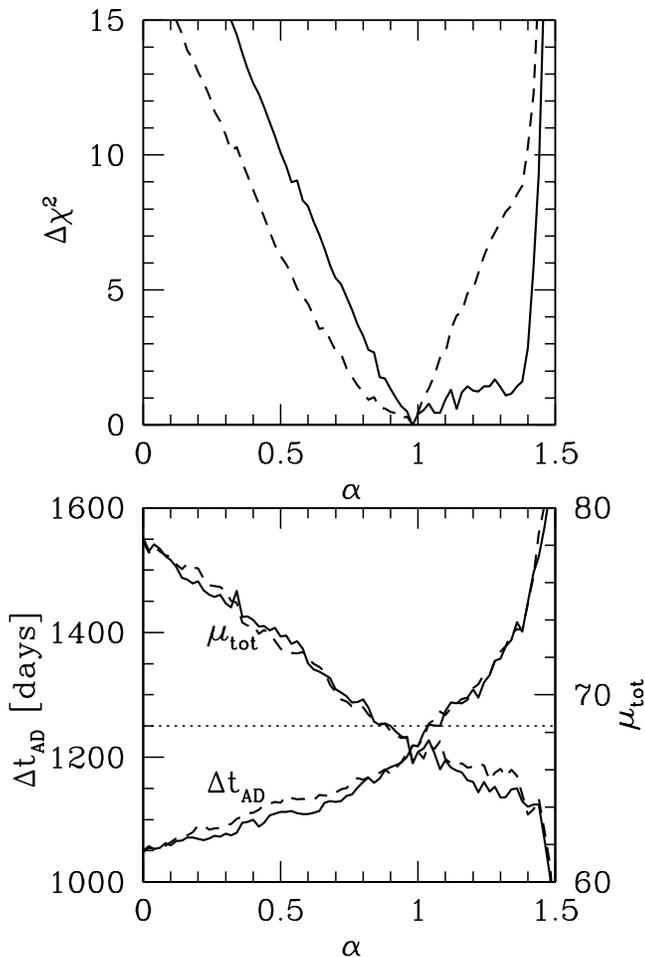}
  \end{center}
  \caption{
 {\it Upper:} The $\chi^2$ difference, 
$\Delta\chi^2=\chi^2-\chi^2({\rm min})$, as a function of the inner
 slope of the dark halo component. For each $\alpha$, the other model
 parameters are optimized to minimize $\chi^2$. The solid line shows
 the default case, whereas the dashed line indicates the result when
 an additional prior of $c_{-2}=6.0\pm1.5$ from the {\it Chandra}
 X-ray observation is included for mass modeling. {\it Lower:}  The
 total magnification factor for the five quasar images, $\mu_{\rm
 tot}$, and the time delay between the quasar images A and D, $\Delta
 t_{\rm AD}$, predicted by the best-fit models as a function of the
 inner dark halo slope $\alpha$. The horizontal dotted line indicates
 the observed lower limit of $\Delta t_{\rm AD}$ reported by
 \citet{fohlmeister08}.}
\label{fig:gnfw_alpha}
\end{figure}

\subsection{Generalized NFW Profile}

Now we allow the inner slope of the dark halo component $\alpha$ to
vary to see how well the current strong lens data, including time
delays which are quite helpful to break degeneracy in mass models
(e.g., \cite{kawano06}), can constrain the inner density profile.  
Specifically, for each fixed value of $\alpha$ we optimize the other
parameters. Figure~\ref{fig:gnfw_alpha} shows the $\chi^2$ difference
as a function of $\alpha$. We find that our mass modeling is quite
consistent with the NFW profile, i.e., $\alpha=1$. We constrain  the
range of the slope to $0.76<\alpha<1.41$ at 95\% confidence limit. The
profile as steep as $\alpha=1.5$ is clearly rejected. There is a clear
correlation between $\alpha$ and the velocity dispersion of galaxy G1
such that the best-fit velocity dispersion decreases with increasing
$\alpha$, which approximately conserves the central core mass of the
total matter distribution. 

As discussed in \S~\ref{sec:xray}, the {\it Chandra} X-ray observation
suggests that the lensing cluster may have larger value of the
concentration parameter than the best-fit NFW model predicts. We
include this effect by adding a prior of $c_{-2}=6\pm 1.5$ to the mass
model to see how the constraint on $\alpha$ is modified. The result
shown in Figure~\ref{fig:gnfw_alpha} indicates that the constraint is
basically shifted to the lower $\alpha$, i.e., shallower inner density
slope. The resulting range is $0.62<\alpha<1.14$ at 95\% confidence
limit.

In Figure~\ref{fig:gnfw_alpha}, we also show the total magnification
factor for the five quasar images, $\mu_{\rm tot}$, and the time delay
between the quasar images A and D, $\Delta t_{\rm AD}$, predicted by
the best-fit model for each fixed $\alpha$. We find that $\mu_{\rm
  tot}$ is decreasing and $\Delta t_{\rm AD}$ is increasing with
increasing $\alpha$, which are consistent with well-known dependence
of the magnification and time delay on the radial density slope (e.g.,
\cite{wambsganss94}; \cite{oguri03}).
Thus the reported lower limit of  $\Delta t_{\rm AD}>1250$~days
\citep{fohlmeister08} prefers a steeper inner slope (larger $\alpha$),
which appears to be opposed to the effect of the prior from X-ray
discussed above. In either case, the strong dependence of $\Delta
t_{\rm AD}$ on $\alpha$ implies that additional A--D time delay
measurements will greatly help to constrain the mass model further. 

\section{Summary}\label{sec:summary}

We have revisited parametric mass modeling of SDSS~J1004+4112, a
unique quasar--cluster lens system, using a newly developed mass
modeling software. We include several new constraints, including
positions of spectroscopically confirmed multiply imaged galaxies,
time delays between some quasar images, and faint central images. Our 
model comprising of a halo component modeled by the generalized NFW
profile, member galaxies including the brightest cluster galaxy G1,
and perturbation terms, well successfully reproduced all observations
including time delays.  

Unlike earlier claims based on parametric mass modeling, we have found
that the center and orientation of the dark halo component is in good
agreement with those of member galaxies and {\it Chandra} X-ray
observation, implying that the cluster is highly relaxed. The radial
profile from strong lensing is also in excellent agreement with the
mass profile inferred from the X-ray observation. Our mass modeling
prefers the dark halo component with the inner slope close to
$\alpha=1$, being consistent with so-called NFW density
profile. Additional measurement of the time delay between quasar image
A and D will be useful to constrain the mass model further. The
predicted total magnification of $\mu_{\rm tot}=67$ for the NFW
profile is quite large compared with those for typical galaxy-scale
lenses, because of the shallower density profiles for clusters. This
makes the lens system a good site for studying the fine structure of
the quasar through microlensing (\cite{richards04}; \cite{green06};
\cite{lamer06}; \cite{pooley07}) or for studying host galaxies of
distant quasars \citep{ross09}. We note that our result based on
parametric mass modeling is broadly consistent with recent
non-parametric mass modeling by e.g., \citet{liesenborgs09}. 

Our result suggests that the core of the lensing cluster at $z=0.68$ is
highly evolved. Recently, \citet{limousin10} showed that the cluster
MACSJ1423.8+2404 at $z=0.54$ is similarly highly relaxed based on
the comparison of mass, light, and gas distributions. Therefore our
result may point to the fact that relaxed clusters are quite common
already at $z\sim 0.6$. 

It is clear that the lensing cluster SDSS~J1004+4112 is currently one
of the best-studied high-redshift clusters whose inner density profile
is very tightly constrained by strong lensing. Additional constraints
on this cluster with weak lensing, Sunyaev-Zel'dovich effect, and
spectroscopic identifications of member galaxies will be important to
advance our understanding of high-redshift clusters. 

\bigskip

I would like to thank the referee, Marceau Limousin, for useful
comments and suggestion.

\appendix
\section{Lens Modeling Software {\it glafic}}
\label{sec:glafic}

We have developed a comprehensive software package called {\it glafic}
that can be used for a wide variety of analysis for gravitational
lensing. Its features include efficient computations of lensed images
for both point and extended sources, handling of multiple sources, a
wide range of lens potentials, and the implementation of noble
technique for mass modeling. Currently the software can be downloaded
from \texttt{http://www.slac.stanford.edu/\~{}oguri/glafic/}. 

In this code, we adopt the adaptive mesh refinement algorithm described
in \citet{keeton01a} for deriving lensed point source images, although
the code uses rectangular coordinates rather than polar coordinates.
Critical curves are computed using a marching squares technique
\citep{jullo07}. In simulations of extended sources, one can convolve
point spread functions and include a number of galaxies read from a
catalog file, which should also be useful for weak lensing work. For
more details, readers are referred to a manual available at the URL
above.  

\section{Source-plane $\chi^2$ Minimization}
\label{sec:schi}

Strong lens modeling with the standard $\chi^2$ minimization is
sometimes time-consuming, especially when many lens potential
components and images are involved. One way to overcome this problem
is to evaluate $\chi^2$ in the source plane instead of the image
plane. Although the source plane $\chi^2$ involves approximations
(given that observational measurements are always made in the image
plane) and therefore is less accurate than the image plane $\chi^2$,
it allows one to estimate $\chi^2$ without solving the nonlinear lens
equation.  This technique has been adopted by several authors
(e.g., \cite{kayser90}; \cite{kochanek91}; \cite{koopmans98};
\cite{keeton01a}; \cite{smith05}; \cite{bradac05}; \cite{halkola06};
\cite{jullo07}; \cite{sand08}), although the implementations were
quite different for different papers. Here we describe our
implementation and argue the accuracy of the technique.  

For a given source position $\mathbf{u}$, $\chi^2$ is estimated as 
\begin{eqnarray}
\chi^2_{\rm img}&=&\chi^2_{\rm pos}+\chi^2_{\rm flux}+\chi^2_{\Delta t},\label{eq:c2}\\
\chi^2_{\rm pos}&=&\sum_i \frac{(\mathbf{x}_{i,{\rm
      obs}}-\mathbf{x}_i)^2}{\sigma_{x_i}^2},\label{eq:c2_pos}\\
\chi^2_{\rm flux}&=&\sum_i \frac{(m_{i,{\rm obs}}+2.5\log\mu_i-m_0)^2}{\sigma_{m_i}^2},\label{eq:c2_flux}\\
\chi^2_{\Delta t}&=&\sum_i\frac{(\Delta t_{i,{\rm obs}}-\Delta
      t_i-\Delta t_0)^2}{\sigma_{\Delta t_i}^2},\label{eq:c2_dt}
\end{eqnarray}
where $\mathbf{x}_i$, $\mu_i$, $\Delta t_i$ is the position,
magnification, and time delay for the $i$-th image, respectively.  
They are related with the source position through the lens 
equation:  
\begin{equation}
\mathbf{u}=\mathbf{x}_i-\mathbf{\nabla}\phi(\mathbf{x}_i).
\end{equation}
Here $\phi(\mathbf{x})$ is the so-called lens potential. From
the lens equation, the magnification factor $\mu_i$ is computed as   
\begin{equation}
\mu_i=\left| \det (\mathtt{M}_i)\right|,
\end{equation}
with $\mathtt{M}_i$ being the magnification tensor defined by
\begin{equation}
\mathtt{M}_i^{-1}=\frac{\partial \mathbf{u}}{\partial \mathbf{x}_i}.
\label{eq:magtensor}
\end{equation}
Finally the time delay $\Delta t_i$ is 
\begin{equation}
\Delta t_i=\frac{1+z_l}{c}\frac{D_{\rm l}D_{\rm s}}{D_{\rm
    ls}}\left[\frac{(\mathbf{u}-\mathbf{x}_i)^2}{2}-\phi(\mathbf{x}_i)\right]. 
\end{equation}

Computing $\chi^2$ in equation (\ref{eq:c2}) usually requires the
derivation of $\mathbf{x}_i$ for a given $\mathbf{u}$, which is
the most time-consuming part because the lens equation is {\it not}
one-to-one mapping and thus the extensive solution finding in the
image plane is needed. In the source plane $\chi^2$ technique, 
we do not solve the lens equation but just approximate $\chi^2$ as
follows. Assuming $\mathbf{x}_i$ and $\mathbf{x}_{i,{\rm obs}}$ are
close with each other, we can write
\begin{equation}
\mathbf{x}_{i,{\rm obs}}-\mathbf{x}_i \approx \mathtt{M}_i(\mathbf{u}_{i,{\rm
    obs}}-\mathbf{u}),
\end{equation}
where $\mathbf{u}_{i,{\rm obs}}$ is the source position computed from
the observed $i$-th image position:
\begin{equation}
\mathbf{u}_{i,{\rm obs}}=\mathbf{x}_{i,{\rm obs}}-\mathbf{\nabla}\phi(\mathbf{x}_{i,{\rm obs}}),
\end{equation}
and $\mathtt{M}_i$ is estimated at $\mathbf{x}=\mathbf{x}_{i,{\rm obs}}$.
Then equation (\ref{eq:c2_pos}) becomes
\begin{equation}
\chi^2_{\rm pos}\approx \chi^2_{\rm pos,src}=\frac{(\mathbf{u}_{i,{\rm
      obs}}-\mathbf{u})^{\rm T}\mathtt{M}_i^2(\mathbf{u}_{i,{\rm obs}}-\mathbf{u})}{\sigma_{x_i}^2}.
\end{equation}
Similarly, the magnification and time delay are approximated as
\begin{equation}
\mu_i\approx \mu(\mathbf{x}_{i,{\rm obs}})+\mathtt{M}_i(\mathbf{u}-\mathbf{u}_{i,{\rm obs}})\cdot\left.\frac{d\mu}{d\mathbf{x}}\right|_{\mathbf{x}=\mathbf{x}_{i,{\rm obs}}}
\end{equation}
\begin{eqnarray}
\Delta t_i&\approx& \Delta t(\mathbf{x}_{i,{\rm obs}})+\frac{1+z_l}{c}\frac{D_{\rm l}D_{\rm s}}{D_{\rm
    ls}}\nonumber\\
&&\times (\mathbf{u}_{i,{\rm obs}}-\mathbf{x}_{i,{\rm obs}})\cdot(\mathbf{u}-\mathbf{u}_{i,{\rm obs}}).
\end{eqnarray}
In the code, $d\mu/d\mathbf{x}$ is evaluated numerically. 
Inserting these expressions into equations (\ref{eq:c2_flux}) and
(\ref{eq:c2_dt}) yield $\chi^2_{\rm flux,src}$ and $\chi^2_{\Delta t,
  {\rm src}}$, respectively. The source plane $\chi^2$ is just a sum
of these three: 
\begin{equation}
\chi^2_{\rm src}=\chi^2_{\rm pos,src}+\chi^2_{\rm
  flux,src}+\chi^2_{\Delta t,{\rm src}},\label{eq:c2src}
\end{equation}
Note that $m_0$ and $\Delta t_0$ are nuisance parameters whose
best-fit values can easily be derived as
\begin{equation}
m_0=\frac{\sum_i\left[(m_{i,{\rm obs}}+2.5\log\mu_i)/\sigma^2_{m_i}\right]}{\sum_i\left(1/\sigma^2_{m_i}\right)},
\end{equation}
\begin{equation}
\Delta t_0=\frac{\sum_i\left[(\Delta t_{i,{\rm obs}}-\Delta
    t_i)/\sigma^2_{\Delta t_i}\right]}{\sum_i\left(1/\sigma^2_{\Delta t_i}\right)}.
\end{equation}
Also note that one can adopt image fluxes as constraints rather than
magnitudes. The modification for this is quite straightforward (see
also \cite{keeton01a}).

\begin{figure}
  \begin{center}
    \FigureFile(82mm,82mm){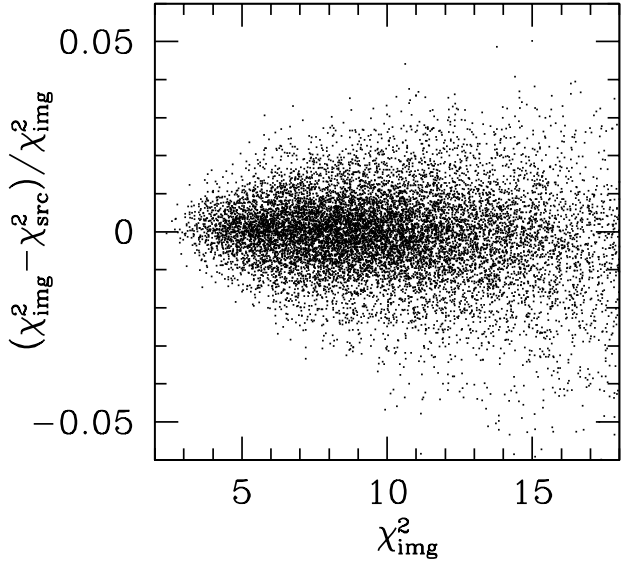}
  \end{center}
  \caption{
  Accuracy of the source plane $\chi^2$, $\chi^2_{\rm src}$. Each
  point shows the fractional difference between $\chi^2_{\rm img}$ and
  $\chi^2_{\rm src}$ as a function of $\chi^2_{\rm img}$, estimated
  at each MCMC step. The assumed mass model is described in the text.
  }
\label{fig:chi2comp}
\end{figure}

We find that the source plane $\chi^2$, if properly evaluated like
above, is accurate enough to be used in most cases of strong lens mass
modeling. As a specific example, we consider a mass model consisting
of an SIE and external shear. The model parameters are $\sigma=320{\rm
  km\,s^{-1}}$, $e=0.3$, $\theta_e=20$~deg, $\gamma=0.1$,
$\theta_\gamma=90$~deg, $z_l=0.5$, $z_s=2.0$, $\mathbf{u}$=(0.04,
$-0.02$). We then add errors to the image positions, magnitudes, time
delays, and fit the ``observed'' four-image system using the same mass
model. We adopt a positional error of $0\farcs01$, a magnitude error
of $0.1$, and a (relative) time delay error of $0.5$~days. We run a
Markov Chain Monte Carlo (MCMC) for this model to explore $\chi^2$
around the best-fit model, and at each step we compute both
$\chi^2_{\rm img}$ and $\chi^2_{\rm src}$ to see the difference of
these. The result shown in Figure~\ref{fig:chi2comp} indicates that
$\chi^2_{\rm src}$ agrees well with $\chi^2_{\rm img}$ within a few
percent level, which are sufficiently accurate to derive the best-fit
mass model and errors on best-fit parameters. We note that
$\chi^2_{\rm src}$ is even more accurate when observational
constraints are tighter. 

We note that \citet{halkola06} also studied accuracy of the source
plane $\chi^2$ based on modeling of the strong lens cluster A1689. 
They found a clear correlation between $\chi^2_{\rm img}$ and
$\chi^2_{\rm src}$, but with different values. This is because they
did not take account of the full magnification tensor (eq. 
[\ref{eq:magtensor}]) but simply adopted a magnification factor to
approximate $\chi^2$, as has been often done in the literature. Thus
our result highlights the quantitative importance of the proper
approximation taking the full lens mapping into account.


\begin{thebibliography}{}

\bibitem[Brada{\v c} et al.(2005)]{bradac05}
Brada{\v c}, M., Schneider, P., Lombardi, M., \& Erben, T.\ 
2005, \aap, 437, 39 

\bibitem[Congdon \& Keeton(2005)]{congdon05}
Congdon, A.~B., \& Keeton, C.~R.\ 2005, \mnras, 364, 1459 

\bibitem[El{\'{\i}}asd{\'o}ttir et al.(2007)]{eliasdottir07} 
El{\'{\i}}asd{\'o}ttir, {\'A}., et al.\ 2007, arXiv:0710.5636 

\bibitem[Evans \& Witt(2003)]{evans03}
Evans, N.~W., \& Witt, H.~J.\ 2003, \mnras, 345, 1351 

\bibitem[Fohlmeister et al.(2007)]{fohlmeister07}
Fohlmeister, J., et al. 2007, \apj, 662, 62 
  
\bibitem[Fohlmeister et al.(2008)]{fohlmeister08}
Fohlmeister, J., Kochanek, C.~S., Falco, E.~E., Morgan, C.~W., 
\& Wambsganss, J. 2008, \apj, 676, 761 
  
\bibitem[Gavazzi(2005)]{gavazzi05}
Gavazzi, R.\ 2005, \aap, 443, 793 

\bibitem[Green(2006)]{green06}
Green, P.~J.\ 2006, \apj, 644, 733 

\bibitem[Halkola et al.(2006)]{halkola06}
Halkola, A., Seitz, S., \& Pannella, M.\ 2006, \mnras, 372, 1425 

\bibitem[Inada et al.(2003)]{inada03}
Inada, N., et al. 2003, \nat, 426, 810 
  
\bibitem[Inada et al.(2005)]{inada05}
Inada, N., et al. 2005, \pasj, 57, L7 
  
\bibitem[Inada et al.(2006)]{inada06}
Inada, N., et al.\ 2006, \apjl, 653, L97 

\bibitem[Inada et al.(2008)]{inada08}
Inada, N., Oguri, M., Falco, E.~E., Broadhurst, T.~J., 
Ofek, E.~O., Kochanek, C.~S., Sharon, K., 
\& Smith, G.~P.\ 2008, \pasj, 60, L27 

\bibitem[Jing, Suto(2000)]{jing00}
Jing, Y.~P., \& Suto, Y. 2000, \apj, 529, L69

\bibitem[Jullo et al.(2007)]{jullo07}
Jullo, E., Kneib, J.-P., Limousin, M., El{\'{\i}}asd{\'o}ttir, 
{\'A}., Marshall, P.~J., \& Verdugo, T.\ 2007, New Journal of Physics,
9, 447  

\bibitem[Kawano et al.(2004)]{kawano04}
Kawano, Y., Oguri, M., Matsubara, T., \& Ikeuchi, S.\ 2004, 
\pasj, 56, 253 

\bibitem[Kawano \& Oguri(2006)]{kawano06}
Kawano, Y., \& Oguri, M.\ 2006, \pasj, 58, 271 

\bibitem[Kayser(1990)]{kayser90}
Kayser, R.\ 1990, \apj, 357, 309 

\bibitem[Keeton(2001a)]{keeton01a}
Keeton, C.~R.\ 2001a, preprint (astro-ph/0102340)

\bibitem[Keeton(2001b)]{keeton01b}
Keeton, C.~R.\ 2001b, preprint (astro-ph/0102341)

\bibitem[Keeton et al.(1997)]{keeton97}
Keeton, C.~R., Kochanek, C.~S., \& Seljak, U.\ 1997, \apj, 
482, 604 

\bibitem[Kochanek(1991)]{kochanek91}
Kochanek, C.~S.\ 1991, \apj, 373, 354 

\bibitem[Koopmans et al.(1998)]{koopmans98}
Koopmans L.~V.~E., de Bruyn A.~G., Jackson N., 1998, MNRAS, 295, 534 

\bibitem[Lamer et al.(2006)]{lamer06}
Lamer, G., Schwope, A., Wisotzki, L., \& Christensen, L.\ 
2006, \aap, 454, 493 

\bibitem[Liesenborgs et al.(2009)]{liesenborgs09}
Liesenborgs, J., de Rijcke, S., Dejonghe, H., \& 
Bekaert, P.\ 2009, \mnras, 397, 341 

\bibitem[Limousin et al.(2010)]{limousin10}
Limousin, M., et al.\ 2010, \mnras, in press (arXiv:0911.4125)

\bibitem[Natarajan et al.(2009)]{natarajan09}
Natarajan, P., Kneib, J.-P., Smail, I., Treu, T., Ellis, R., 
Moran, S., Limousin, M., \& Czoske, O.\ 2009, \apj, 693, 970 

\bibitem[Navarro et al.(1997)]{navarro97}
Navarro, J.~F., Frenk, C.~S., \& White, S.~D.~M.\ 1997, 
\apj, 490, 493 
  
\bibitem[Oguri \& Kawano(2003)]{oguri03}
Oguri, M., \& Kawano, Y.\ 2003, \mnras, 338, L25 

\bibitem[Oguri et al.(2004)]{oguri04}
Oguri, M., et al. 2004, \apj, 605, 78 

\bibitem[Oguri et al.(2005)]{oguri05}
Oguri, M., Takada, M., Umetsu, K., \& 
Broadhurst, T.\ 2005, \apj, 632, 841 

\bibitem[Oguri et al.(2008)]{oguri08}
Oguri, M., et al.\ 2008, \apjl, 676, L1 

\bibitem[Ota et al.(2006)]{ota06}
Ota, N., et al. 2006, \apj, 647, 215 

\bibitem[Pooley et al.(2007)]{pooley07}
Pooley, D., Blackburne, J.~A., Rappaport, S., \& 
Schechter, P.~L.\ 2007, \apj, 661, 19 

\bibitem[Richards et al.(2004)]{richards04}
Richards, G.~T., et al.\ 2004, \apj, 610, 679 

\bibitem[Ross et al.(2009)]{ross09}
Ross, N.~R., Assef, R.~J., Kochanek, C.~S., Falco, E., \& 
Poindexter, S.~D.\ 2009, \apj, 702, 472 

\bibitem[Saha et al.(2006)]{saha06}
Saha, P., Read, J.~I., \& Williams, L.~L.~R.\ 2006, \apjl, 652, L5 

\bibitem[Saha et al.(2007)]{saha07}
Saha, P., Williams, L.~L.~R., \& Ferreras, I.\ 2007, \apj, 663, 29 

\bibitem[Sand et al.(2008)]{sand08}
Sand, D.~J., Treu, T., Ellis, R.~S., Smith, G.~P., \& 
Kneib, J.-P.\ 2008, \apj, 674, 711 

\bibitem[Sharon et al.(2005)]{sharon05}
Sharon, K., et al. 2005, \apjl, 629, L73 
  
\bibitem[Smith et al.(2005)]{smith05}
Smith, G.~P., Kneib, J.-P., Smail, I., Mazzotta, P., Ebeling, H., 
\& Czoske, O.\ 2005, \mnras, 359, 417 

\bibitem[Wambsganss \& Paczynski(1994)]{wambsganss94}
Wambsganss, J., \& Paczynski, B.\ 1994, \aj, 108, 1156 

\bibitem[Williams \& Saha(2004)]{williams04}
Williams, L.~L.~R., \& Saha, P.\ 2004, \aj, 128, 2631 

\bibitem[Yoo et al.(2006)]{yoo06}
Yoo, J., Kochanek, C.~S., Falco, E.~E., \& McLeod, 
B.~A.\ 2006, \apj, 642, 22 

\end{thebibliography}
\end{document}